\documentclass{PoS}

\newcommand{\mh}{M_{\rm H}}
\newcommand{\mw}{M_{\rm W}}

\title{Primordial magnetic fields at preheating.}

\ShortTitle{Primordial magnetic fields at preheating.}
\author{\speaker{A. D\'{\i}az-Gil}%
         \thanks{Supported by FPU grant of MEC.}\\
Departamento de F\'{\i}sica Te\'orica\\
Universidad Aut\'onoma de Madrid, Cantoblanco, 28049 Madrid, Spain\\
\email{andres.diazgil@uam.es}}

\author{J. Garc\'{\i}a-Bellido\\
Departamento de F\'{\i}sica Te\'orica and Instituto de F\'{\i}sica Te\'orica
UAM-CSIC\\
Universidad Aut\'onoma de Madrid, Cantoblanco, 28049 Madrid, Spain\\
\email{juan.garciabellido@uam.es}}
\author{ M. Garc\'{\i}a P\'erez\\
Instituto de F\'{\i}sica Te\'orica UAM-CSIC\\
Universidad Aut\'onoma de Madrid, Cantoblanco, 28049 Madrid, Spain\\
\email{margarita.garcia@uam.es}}

\author{A. Gonz\'alez-Arroyo\\
Departamento de F\'{\i}sica Te\'orica and Instituto de F\'{\i}sica Te\'orica 
UAM-CSIC\\
Universidad Aut\'onoma de Madrid, Cantoblanco, 28049 Madrid, Spain\\
\email{antonio.gonzalez-arroyo@uam.es}}
\abstract{Using lattice techniques we investigate the  
generation of long range cosmological magnetic fields during a cold 
electroweak transition. We will show how magnetic fields arise, during 
bubble collisions, in the form of magnetic strings. We conjecture that
these magnetic strings originate from the alignment of magnetic dipoles 
associated with EW sphaleron-like configurations. We also discuss the early 
thermalisation of photons and the turbulent behaviour of the scalar fields 
after tachyonic preheating.}

\FullConference{The XXV International Symposium on Lattice Field Theory\\
		 July 30 - August 4 2007\\
		 Regensburg, Germany}

\begin{document}

\section{Introduction.}
There have been many theoretical attempts to explain the origin of
large scale cosmological magnetic fields (LSMF)~\cite{giova}.  The
main difficulty resides in understanding their correlation scale which
ranges from the size of galaxies to clusters and super-clusters with
an amplitude of the order of micro-gauss, pointing to a primordial origin.  
Following the work initiated in~\cite{lat05}, we address this issue 
in the context of a cold electroweak transition taking place after a 
period of hybrid inflation. The EW transition has been in the heart of 
many proposals to address magnetogenesis, linking it in many cases with the
generation of the baryon asymmetry. The results presented here
resemble the mechanism proposed by Vachaspati connecting the appearance
of magnetic fields to that of sphalerons and Z-strings~\cite{vachas}.

The model we have considered is a hybrid inflation model with the bosonic 
field content of the Standard Model coupled, via the Higgs field, to a 
singlet inflaton:
\begin{eqnarray}
{\cal L} = - \frac{1}{4}G^a_{\mu\nu}G^{\mu\nu}_a - \frac{1}{4}F^{Y}_{\mu\nu}F_{Y}^{\mu\nu}+ {\rm Tr}\Big[(D_\mu\Phi)^\dag D^\mu\Phi\Big ]+ \frac{1}{2}(\partial_\mu\chi)^2 - V(\Phi,\chi)\\
{\rm V} (\Phi,\chi) =
{\rm V}_0 + \frac{1}{2}(g^2\chi^2-m^2)\,|\Phi|^2 + \frac{\lambda}{4}
|\Phi|^4 + \frac{1}{2} \mu^2 \chi^2 \,
\end{eqnarray}
The couplings are fixed to the standard model values for several
ratios of the Higgs to W masses. For concreteness, we have fixed the
inflaton to Higgs coupling by the relation: $g^2= 2\lambda $. 
To solve the time evolution of the system, starting at the end of
inflation, we have performed a numerical evolution based on a suitable
classical approximation (more details can be found in
Refs.~\cite{marga2,lat05,articulo}). In this work we discuss the
results for $\mh = 4.65\ \mw$.  Results for more realistic values will
be presented in~\cite{articulo}.

\section{Simulation results.}

The evolution of the scalars shows two regimes that determine the
behaviour of the system.  The first one is the spontaneous symmetry
breaking (SSB) stage which proceeds via the generation, expansion and
collision of bubble-like structures in the Higgs field~\cite{marga2}.
It is in this regime where a rich phenomenology can arise, including
the generation of the baryon asymmetry of the Universe and a
stochastic background of gravitational waves, see
Refs.~\cite{marga2,bellido,smit,smitfam}, and where we place the
origin of the LSMF. In our simulations this regime starts right after
the end of inflation ($mt\!\sim\! 5\!-\!10$ for $\mh = 4.65\ \mw$) and
lasts until times of order $mt \!\sim \!35\!-\!40$. After this
relatively short stage, a period of slow approach to equilibrium
follows where two important phenomena appear: turbulence and
thermalisation of high momentum photons.

\subsection{First regime: Generation of magnetic fields.}

\begin{figure}
\centerline{
\hspace*{-.2cm}\includegraphics[width=8.3cm]{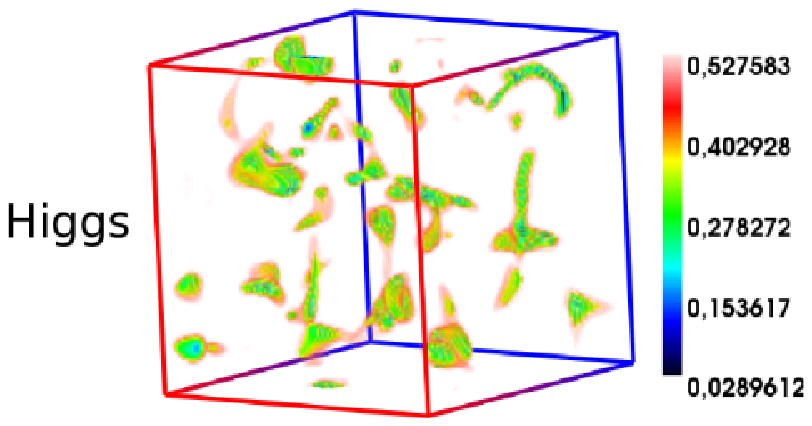}
\hspace*{-.2cm}\includegraphics[width=8.3cm]{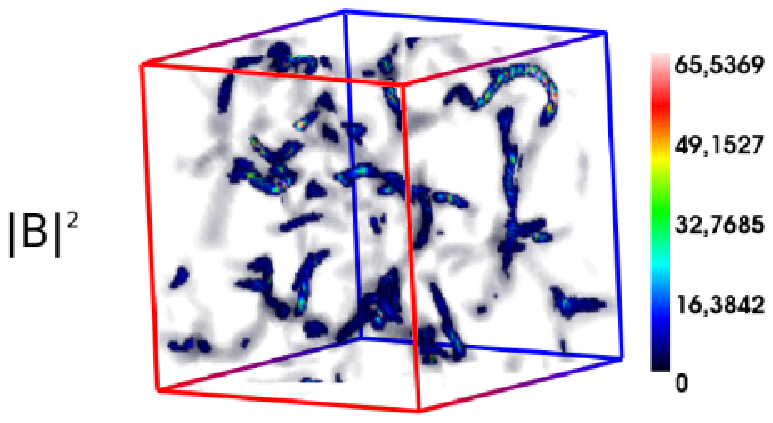}} 
\centerline{
\hspace*{-.2cm}\includegraphics[width=8.3cm]{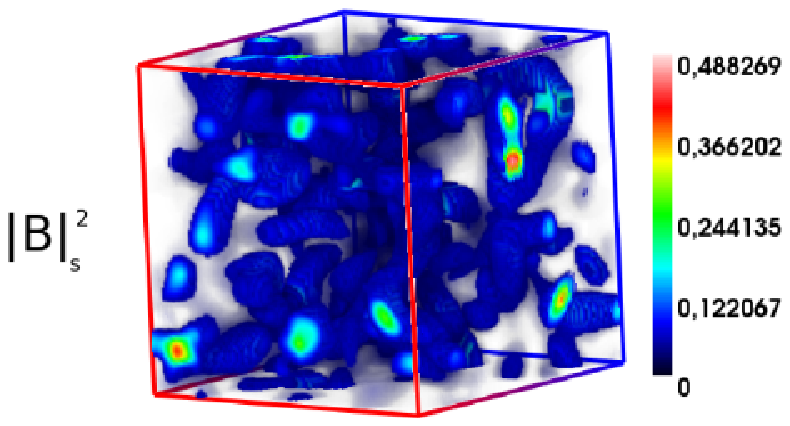}
\hspace*{-.2cm}\includegraphics[width=8.3cm]{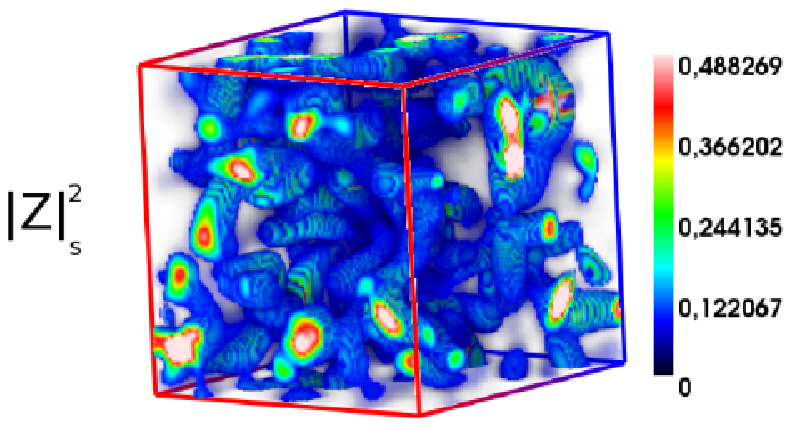}}
\caption{Top: (Left) Locus of points where the value of the Higgs
norm is below $0.53$. (Right) Locus of points where the U(1)-magnetic 
energy is above  $1.5$.  
Bottom: (Left) Smeared density of the U(1)-magnetic energy. 
(Right) Smeared density of the magnetic component of the Z field. Both above 0.1. 
All at  $mt=15$.} 
\label{fig:higgsymag}
\end{figure}

The simulations show that magnetic fields are predominantly generated
in the SSB regime. We have already mentioned that bubble collisions
are responsible for SSB. Bubbles originate at peaks of the initial
Higgs Gaussian random field and expand from there on~\cite{marga2}.
As exhibited in Fig.~\ref{fig:higgsymag} (Left), the regions between
bubbles remain for a longer time in the false vacuum, giving rise to
string-like structures of local minima of the Higgs-field.  Strongly
correlated with them, magnetic fields appear dominantly in the form of
magnetic strings, see Fig.~\ref{fig:higgsymag} (Right). We conjecture
that these magnetic strings originate from sphaleron-like
configurations which are copiously produced during SSB at the location
of Higgs-field zeroes~\cite{marga2,smitfam}.  For non-zero Weinberg
angle the magnetic field generated by the sphaleron looks like a
magnetic dipole~\cite{manton,hind}.  Then, as in a ferromagnetic
material, these dipoles would tend to align along the string of
Higgs-zeroes generating a magnetic string.  One remark has to be done
concerning Fig.~\ref{fig:higgsymag}.  The plot has a lower cutoff
giving the impression that magnetic lines are open. This is just a
projection effect, and places where the magnetic strings seem to end
are just places where the magnetic flux spreads.

The picture that we extract resembles the mechanism proposed in
Ref.~\cite{vachas}, in which magnetic strings also appear at the
electroweak phase transition, and are related to Z-strings. The bottom
half of Fig.~\ref{fig:higgsymag} shows the smeared Z-magnetic field
together with the smeared U(1)-magnetic field at $mt=~15$. The
correlation is manifest.  In this plot the string structure has been
enhanced by performing smearing (averaging the fields over the
neighborhood of a point).  This helps in two ways: first, it smooths
the central singularities of the strings and secondly, it clears up the
radiation background of non-coherent fields. The procedure respects
the regions where the scales are roughly constant and does not disturb
the whole picture as far as the neighborhood involved in the averaging
is smaller than the width of the string ($\sim 1/\mw$).


Having a mechanism for generating magnetic fields, two important questions 
arise, and the viability of this mechanism as LSMF generator depends
on their answer.  These questions are:
\begin{itemize}
\item
Are the scales of the generated magnetic fields large enough to be the seed 
of LSMF? 
\item
Do they persist in time long enough to be used as seeds of LSMF?
\end{itemize}

\begin{figure}
\centerline{
\includegraphics[width=5.5cm,angle=-90]{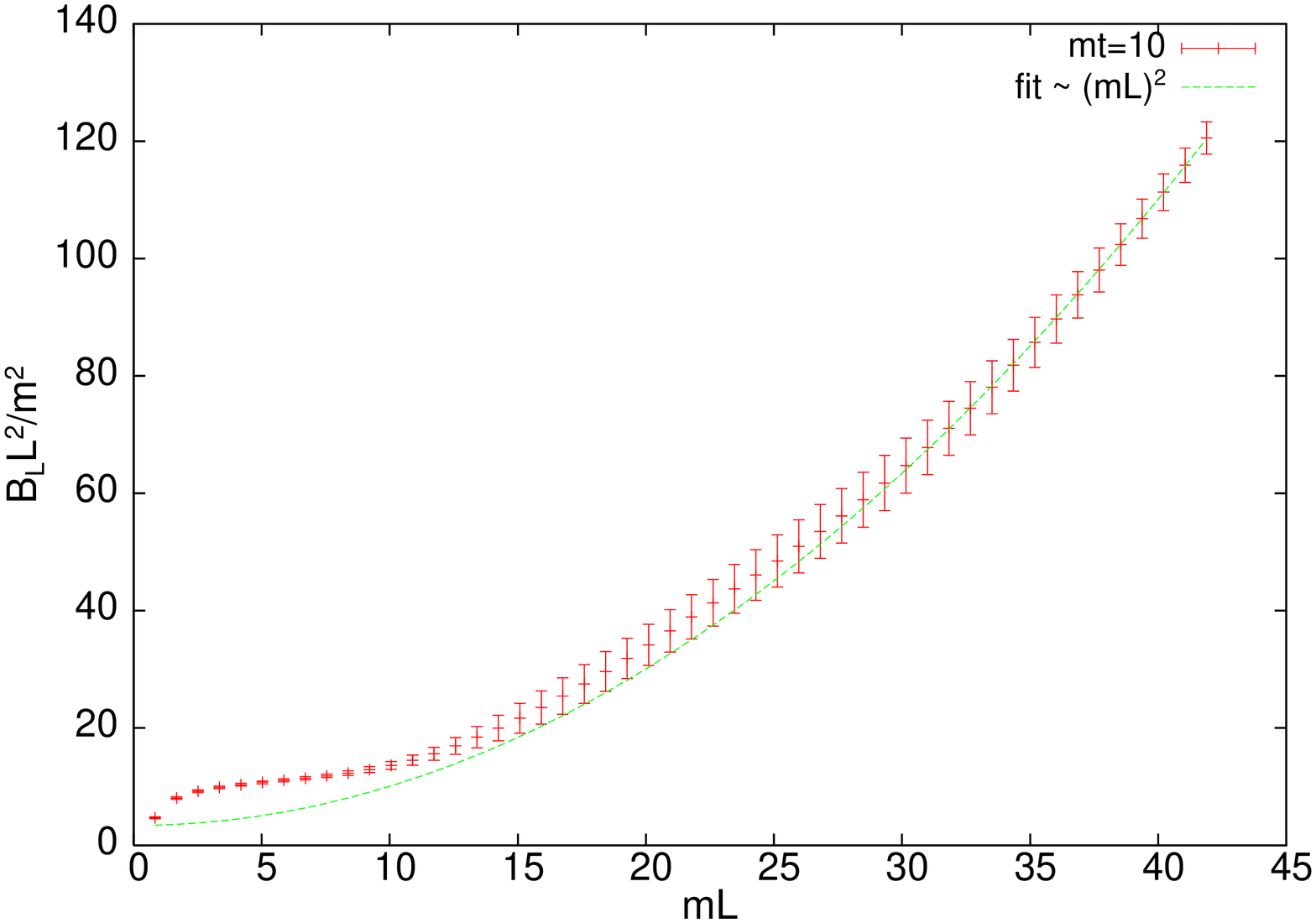}
\includegraphics[width=5.5cm,angle=-90]{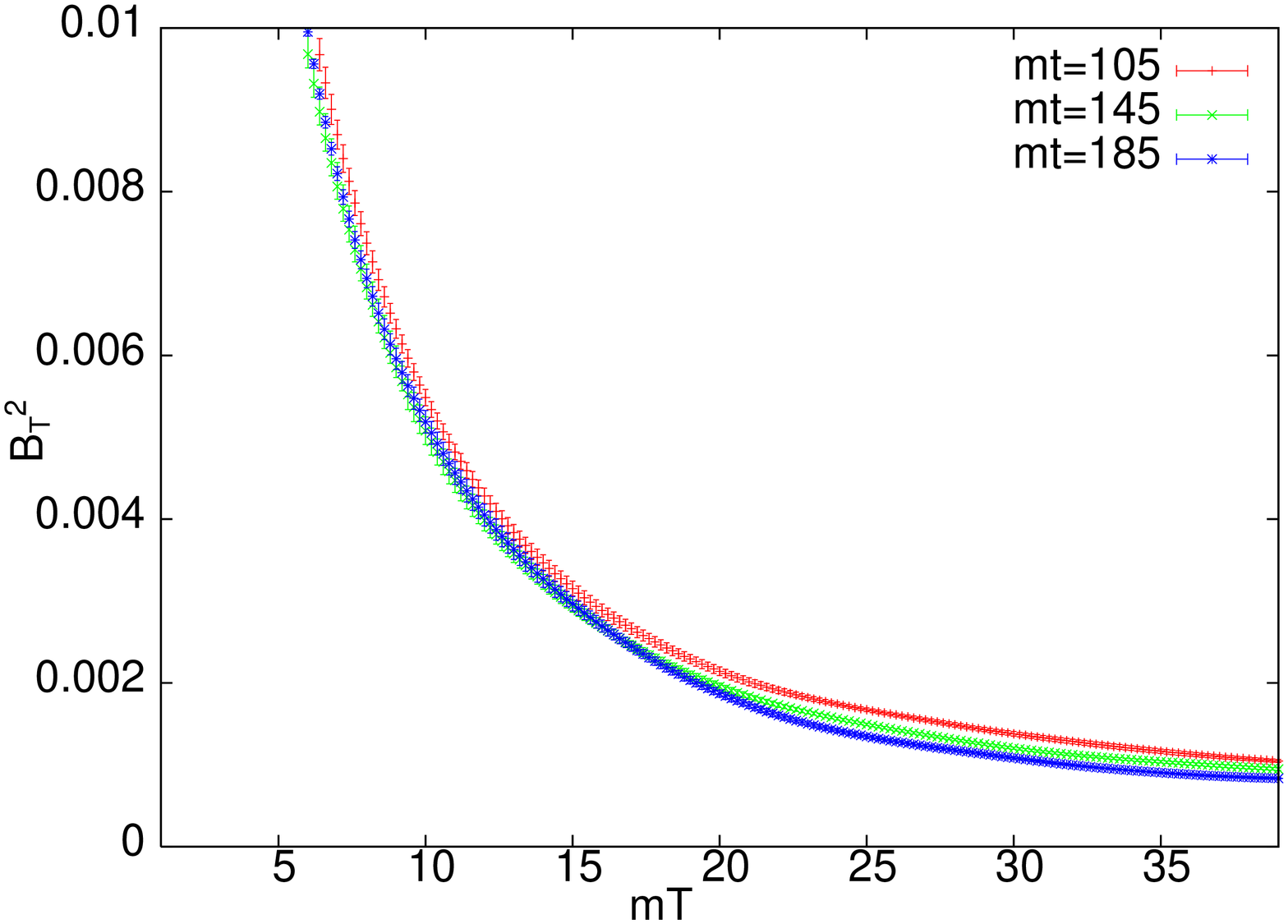}}
\caption{Left: Averaged $B_L(r_0)$, Eq. (2.1), at $mt=10$.  The center
of the box is chosen to be the maximum value in the center of a
string. The plateau length provides an estimate of the minimal length
of a representative string.  Right: Dependence of $B_T^2 \equiv 
\langle \vec B_T^2 \rangle$ on the time-interval length $T$, see Eq. (2.2). 
$B_T^2$ is normalized to the total energy density.}
\label{fig:avplato}
\end{figure}

A qualitative answer to the first question can be extracted from
Fig.~\ref{fig:higgsymag}. Individual strings merge due to magnetic
reconnection forming a cluster of the order of the box size. This
merging of individual strings is essential to enhance the large scale
structure of the fields.  In order to give a quantitative measure of
the nature of the magnetic field objects as actual strings, we have
analysed the following quantity:
\begin{equation}
B_L(r_0)= \frac{1}{L^3}\int_{L(r_0)} dx^3 B^2(x)\,,
\label{eq:bav}
\end{equation}
where the integration is on a box of length $L$, centered at a point
$r_0$ belonging to one of the strings. Figure~\ref{fig:avplato} (Left)
shows the $L$ dependence of $L^2 B_L(r_0)$, averaged over several
configurations.  The expected behaviour of $B_L(r_0)$ if $r_0$ is in
the central core of a string-like object is clear. Inside the string
the magnetic intensity is roughly constant, so the integral goes like
the volume of the box and $B_L(r_0)$ is constant. Once the size of the
box gets larger than the diameter of the string, the integral only
grows along the direction of the string and $B_L(r_0) \sim L^{-2}$.
This gives the plateau seen in the figure.  The small errors indicate
that the string objects are quite configuration independent.The length
of the plateau provides an estimate of the minimal length of the
string as an isolated object. It indicates a mean minimal string
length of order $ \mw L \sim 4-5$, which is a significant part of the
total length. For larger values of $L$ the merging of strings and the
fact that more strings are entering the box, change the behaviour to a
volume like one, with $B_L(r_0)$ roughly constant.

The time persistence analysis is quite more subtle. A detailed analysis of
the behaviour in time of the low momentum magnetic field spectra is on the
way and will be presented in~\cite{articulo}. Some preliminary conclusions 
can already be presented here. A time average can be defined as follows:
\begin{equation}
 \vec B_T(t_0)=\frac{1}{T}\int_{T(t_0)} dt \vec B(t),
\label{eq:bavt}
\end{equation}
this is, the integral over a time interval of length $T$ that starts at $t_0$. 
This time-averaging can be seen as a way to eliminate
short scale magnetic fields, as the radiation fields. For instance,
for a monochromatic wave, 
\begin{eqnarray}
\langle |\vec B_T|^2\rangle \sim \frac{1}{V}\int dk|\vec B_k|^2 \ W^2(k,T)\;,\;W=\frac{sin(|k|T/2)}{|k|T/2},
\label{eq:timeav}
\end{eqnarray}
where the average is both over space and configurations.  For large
$T$, momenta with $|k|>>(T)^{-1}$ are suppressed.  We can use that to
estimate the length of the time interval for which all momenta but our
minimal momenta $p_{\rm min}= 2\pi /L$ are suppressed.  This gives a
window of possible intensities of the magnetic seeds coherent at
scales larger or equal to the scale given by the size of the box.
Fig.~\ref{fig:avplato} (Right) shows the dependence of $\langle |\vec
B_T|^2 \rangle$, normalized to the total energy density ($\rho_0$),
on the length of the time interval $T$, for $p_{\rm min}=0.15m$.
Note that the average turns out to be quite independent of the starting 
point of the averaging $t_0$, indicating that the sub-leading large scales 
are quite independent of time.
From the values above $mT\sim 6$ we obtain a window $\langle\vec B
_T^2 \rangle\sim [10^{-2},10^{-3}] \ \rho_0$.  Assuming the magnetic
field expands as radiation, this would give magnetic fields today of
order $[1,0.3]\mu G$, which are in the range of the observed ones in
galaxies, and even in clusters of galaxies, where no-extra
amplification through a dynamo mechanisms is expected.

\begin{figure}
\centerline{
\includegraphics[width=6.5cm,angle=-90]{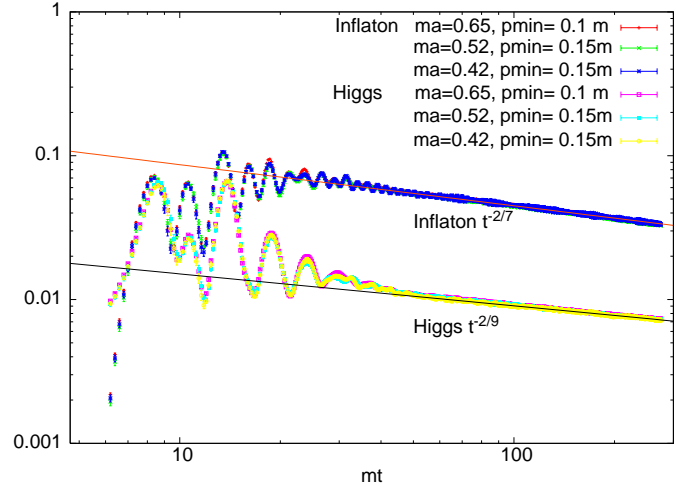}}
\caption{Scalar variances, Eq. (2.4), plotted in logarithmic scale for 
different lattice spacings and volumes.}
\label{fig:turbu}
\end{figure}

\subsection{Second regime: Turbulence and photon thermalisation.}

\begin{figure}
\centerline{
\includegraphics[width=5.5cm,angle=-90]{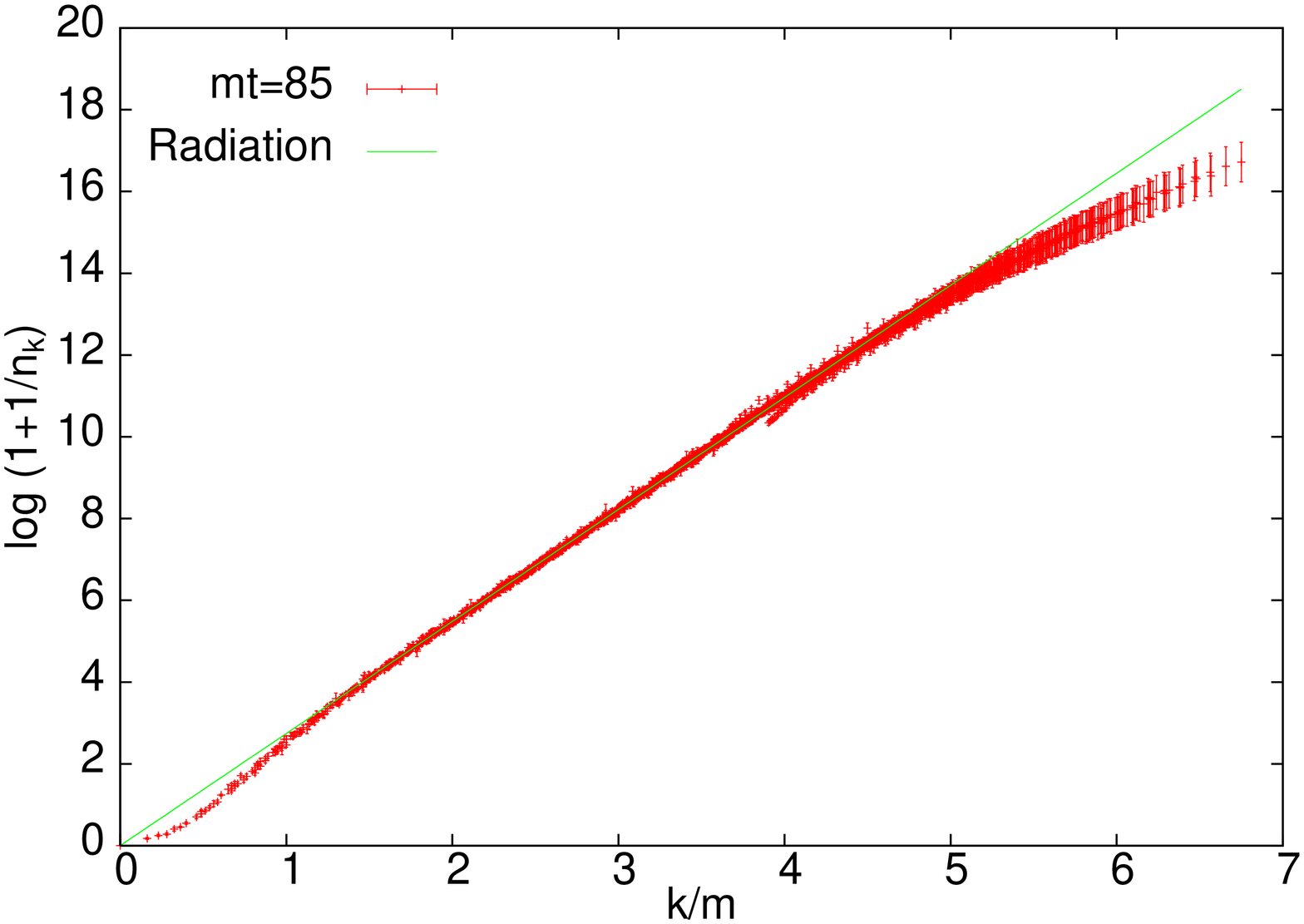}
\includegraphics[width=5.5cm,angle=-90]{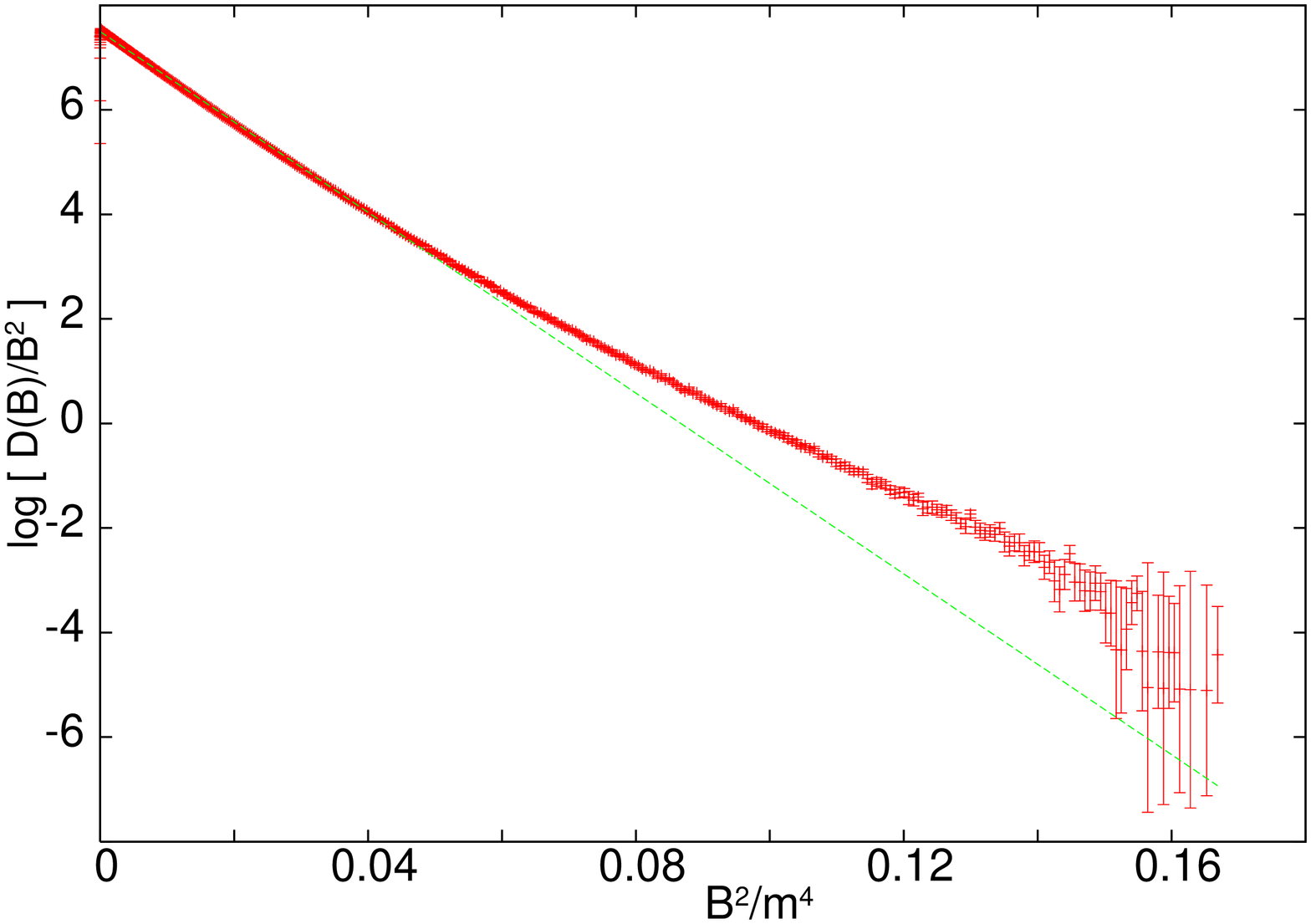}}
\caption{Left: $\rm{log}(n_k^{-1}+1)$ and Right:
$\rm{log}(D(B)/B^2)$, Eq. (2.5), for the spectrum of 
photon radiation at times  $mt=85$ and $mt=225$ respectively.}
\label{fig:spec}
\end{figure}

Once the SSB takes place, the system starts a slow approach to equilibrium.
This proceeds via a period in which the scalars and also some of the gauge 
degrees of freedom undergo a turbulent regime.
Turbulence in the scalar fields manifests through the time evolution
of their variances shown in Fig.~\ref{fig:turbu}. 
As predicted by~\cite{MyT} and also shown in~\cite{lat05,GW}, variances
follow the law:
\begin{equation}
\langle f^2\rangle - \langle f\rangle^2 \sim t^{-\nu}\,; \hspace{5mm}
\nu = {2 \over (2m-1)},
\end{equation}
for $m$-particle interactions and $f$ denoting each of the scalars.
The change in model parameters with respect to~\cite{lat05}, where
$\mw = 6$ GeV, results in a modification of the turbulent behaviour.
The parameter $m$, now turns out to be $m=5 (4)$ for the Higgs
(inflaton), which differs from the former $m=4 (3)$ obtained
in~\cite{lat05}, but is in agreement with~\cite{GW}.

In Ref.~\cite{lat05} we showed the self-similarity of the momentum
spectra of the SU(2) degrees of freedom long after SSB. This
additional sign of turbulence indicated a departure from thermal
equilibrium.  High momentum photons, however, thermalise very early in
the evolution.  In thermal equilibrium the radiation occupation number
is expected to behave according to the Bose-Einstein distribution:
$n_k= (\exp(\beta k) -1 )^{-1}$, from where the thermalisation
temperature can be extracted.  Fig.~\ref{fig:spec} (Left) represents
$\ln(1+n_k^{-1})$ for photon radiation.  Photons with momenta
above $m$ follow the Bose-Einstein law.  The temperature extracted is
of the order of $\mw$ from times $mt \sim 40$ on.  Thermalisation is
corroborated by analysing the distribution of norms of the magnetic
field. In thermal equilibrium this distribution is given by
\cite{articulo}:
\begin{eqnarray}
 D(B)= B^2 \ e^{ -3 B^2 / (2 \langle B^2 \rangle)}\,; \hspace{5mm}
\langle B^2 \rangle =  \frac{\pi^2 T^4}{15}
\end{eqnarray}
Fig.~\ref{fig:spec} right shows $D(B)$. From time $mt \sim 40$ on the 
equilibrium distribution is reached,
with a temperature in agreement with the one extracted from the spectra. 
In both plots a systematic deviation from the thermal behaviour is
observed at the low momenta, large norm, part of the fits. It is
there where the interesting magnetic string structures reside. An analysis of 
the possible, late time, turbulent and helical behaviour of the low momentum 
part of the magnetic field spectrum is on the way and will be reported elsewhere.

\section{Conclusions.}

We have analysed numerically the proposal that long range magnetic
fields could be generated during a cold electroweak transition after a
period of low scale hybrid inflation.  The generation mechanism is
mainly based on two facts:

\begin{itemize}

\item At the SSB stage bubble-like structures, associated to
local maxima in the Higgs-field norm, appear. Points outside the
bubble front, remaining close to the false vacuum, form string like
structures.

\item Bubble collisions give rise to sphaleron-like
configurations attached to the location of zeroes of the Higgs field.
For  $\theta_W \ne 0$ these sphalerons behave as magnetic dipoles.

\end{itemize}

These two ingredients together lead to an allignment of the
sphalerons' dipoles forming magnetic string networks as we have
observed in our simulations. Some results concerning the coherence and
intensity of the generated magnetic fields have been presented. A
further analysis of the time evolution of the magnetic field, as well
as the study of the dependence of the $\mh$ to $\mw$ ratio will be
presented in~\cite{articulo}.

We have also discussed some features of the late time behaviour of the
system, among them, thermalisation of photon radiation and turbulence
in the scalar fields.

\acknowledgments
We acknowledge financial support from the  Madrid Regional Government (CAM),
under the program HEPHACOS P-ESP-00346, and the Spanish Research Ministry
(MEC), under contracts FPA2006-05807,  FPA2006-05485, FPA2006-05423.
Also acknowledged is the use of the MareNostrum supercomputer at the BSC-CNS
and the IFT-UAM/CSIC computation cluster.

\end{document}